\documentclass{elsart}

 \usepackage{graphicx}

\usepackage{amssymb}

\begin{document}

\begin{frontmatter}



  \title{Neutron Interactions as Seen by A Segmented Germanium Detector}


  \author{I.~Abt},
  \author{A.~Caldwell},
  \author{K.~Kr\"oninger\thanksref{goet}},
  \author{J.~Liu\corauthref{cor}}\ead{jingliu@mppmu.mpg.de},
  \author{X.~Liu},
  \author{B.~Majorovits}

  \address{Max-Planck-Institut f\"ur Physik, M\"unchen, Germany}

  \corauth[cor]{\textit{Telephone:} +49-(0)89-32354-415}

  \thanks[goet]{\textit{Present address:} II. Physikalisches
    Institut, G\"ottingen, Germany}

  \begin{abstract}
    The GERmanium Detector Array, GERDA, is designed for the search
    for ``neutrinoless double beta decay'' (0$\nu$2$\beta$) with
    germanium detectors enriched in $^{76}$Ge. An 18-fold segmented
    prototype detector for GERDA Phase II was exposed to an AmBe
    neutron source to improve the understanding of neutron induced
    backgrounds. Neutron interactions with the germanium isotopes
    themselves and in the surrounding materials were studied. Segment
    information is used to identify neutron induced peaks in the
    recorded energy spectra.

    The Geant4 based simulation package MaGe is used to simulate the
    experiment. Though many photon peaks from germanium isotopes
    excited by neutrons are correctly described by Geant4, some
    physics processes were identified as being incorrectly treated or
    even missing.
  \end{abstract}

  \begin{keyword}

    double beta decay \sep germanium detector \sep segmentation \sep
    neutron interaction


    \PACS 23.40.-s \sep 14.60Pq \sep 28.20.-v \sep 29.40.-n
  \end{keyword}
\end{frontmatter}

\section{Introduction}
\label{sec:int}
The GERDA (GERmanium Detector Array) experiment~\cite{gerda}, designed
for the search for ``neutrinoless double beta decay''
(0$\nu$2$\beta$), is currently under construction in Hall A of the
INFN Gran Sasso National Laboratory (LNGS), Italy. Neutrons produced
near the germanium detectors by penetrating cosmic-ray muons can
induce background events. In addition, neutrons from $(\alpha, n)$
reactions in the surrounding rock are also a potential source of
background. The study of neutron interactions with germanium isotopes
as well as the surrounding materials is thus of great interest.

Segmented germanium detectors will be used in GERDA Phase II. It has
been shown that segment information is very useful to identify photon
induced background~\cite{pid}. It is interesting to check if segment
information can also help with the identification of neutron induced
background.

In order to study the issues mentioned above, a GERDA Phase II 18-fold
segmented prototype detector~\cite{siegfried} was exposed to an AmBe
neutron source. Energy spectra were recorded for each segment and the
core. The segment information was used to identify peaks induced by
neutron interactions.

The Geant4~\cite{g1,g2} based simulation package, MaGe~\cite{mage},
has been co-developed by the GERDA and MAJORANA~\cite{major}
collaborations. The simulation of neutron interactions was verified by
comparisons to data.

\section{Experimental Setup and Data Sets}
\label{sec:exp}
The detector used is the first segmented GERDA prototype detector. The
true coaxial cylindrical crystal has a height of 70~mm and a diameter
of 75~mm with a 10~mm hole in the center. It is 18-fold segmented with
a 6-fold segmentation in the azimuthal angle $\phi$ and a 3-fold
segmentation in the height $z$. It was operated in a conventional test
cryostat. Details of the detector and its cryostat can be found in
\cite{siegfried}.

A 1.1~GBq isotropic AmBe neutron source was used in the
experiment. The energy spectrum of the neutrons emitted from the
$^{9}$Be$(\alpha,n)^{12}$C$^{*}$ nuclear reaction extends to
12~MeV. High resolution measurements of the neutron energy spectra of
this kind of neutron source are presented in~\cite{amben, geiger}. The
dependence of the emittance of photons from the de-excitation of
$^{12}$C$^{*}$ on the neutron energy is described in~\cite{geiger}.

\begin{figure}[tbhp]
  \centering
  \includegraphics[width=0.8\textwidth]{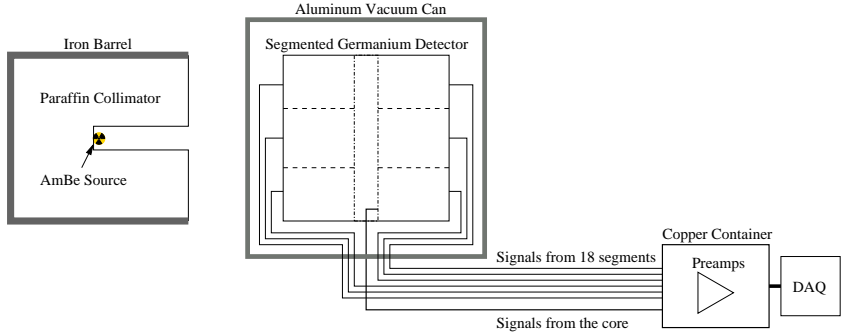}
  \caption{Schematic experimental setup (not to scale).}
  \label{fig:exp}
\end{figure}

The neutron source was located in a cylindrical paraffin
collimator. The schematic experimental setup (not to scale) is shown
in Fig.~\ref{fig:exp}. The center of the collimator was vertically
aligned to the center of the detector and the distance between source
and detector center was about 1~m.

The core and segment electrodes were connected to charge sensitive
pre-amplifiers. Their output was digitized using 14-bit ADCs in an XIA
Pixie-4 data acquisition system~\cite{daq} with a sampling rate of
75~MHz, and recorded separately when the core was triggered. Two
different gain factors were chosen for four different
measurements. The data sets are listed in Tab.~\ref{tab:datset}. A low
gain factor was chosen so that the energy range up to $\sim 11$~MeV
could be covered. A high gain factor was chosen for measurements up to
$\sim 3$~MeV.

Two measurements were performed with the AmBe source present. They are
referred to as HGdat (High Gain data) and LGdat (Low Gain data). In
order to determine the background from the laboratory environment two
more measurements without the source were performed. They are referred
to as HGbg (High Gain background) and LGbg (Low Gain background). The
data samples with different gains were combined for the study below
3~MeV.

\begin{table}[tbhp]
  \caption{Data sets recorded with and without source.} 
  \label{tab:datset}
  \begin{tabular*}{\textwidth}{@{\extracolsep{\fill}}lcccc}\hline\hline
    & \multicolumn{2}{c}{With AmBe Source} & \multicolumn{2}{c}{Without AmBe Source} \\\hline
    DAQ Gain & High Gain & Low Gain  & High Gain & Low Gain \\
    $E_{max}$ [MeV] & $\sim$ 3.5  & $\sim$ 11 & $\sim$ 3.5 & $\sim$ 11 \\
    No. of Events & 7.1 M & 4.7 M & 1.5 M & 4.7 M \\
    Name & HGdat & LGdat & HGbg & LGbg \\\hline\hline
  \end{tabular*}
\end{table}

\section{Core Spectra}
\label{sec:spec}
\begin{figure}[tbhp]
  \centering
  \includegraphics[width=\textwidth,clip]{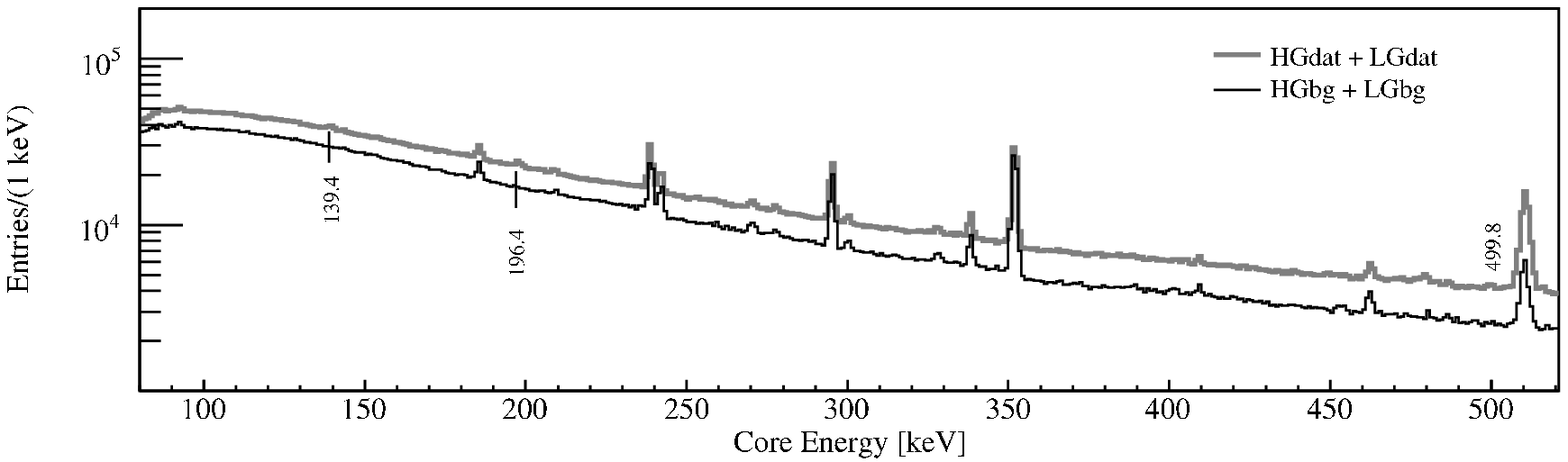}
  \includegraphics[width=\textwidth,clip]{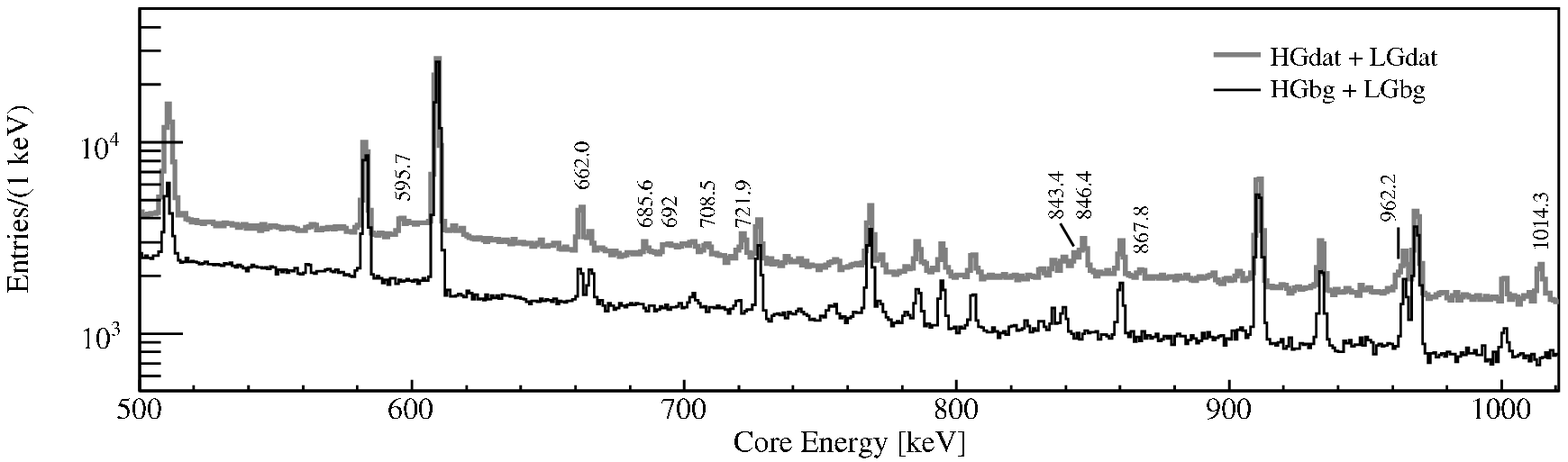}
  \includegraphics[width=\textwidth,clip]{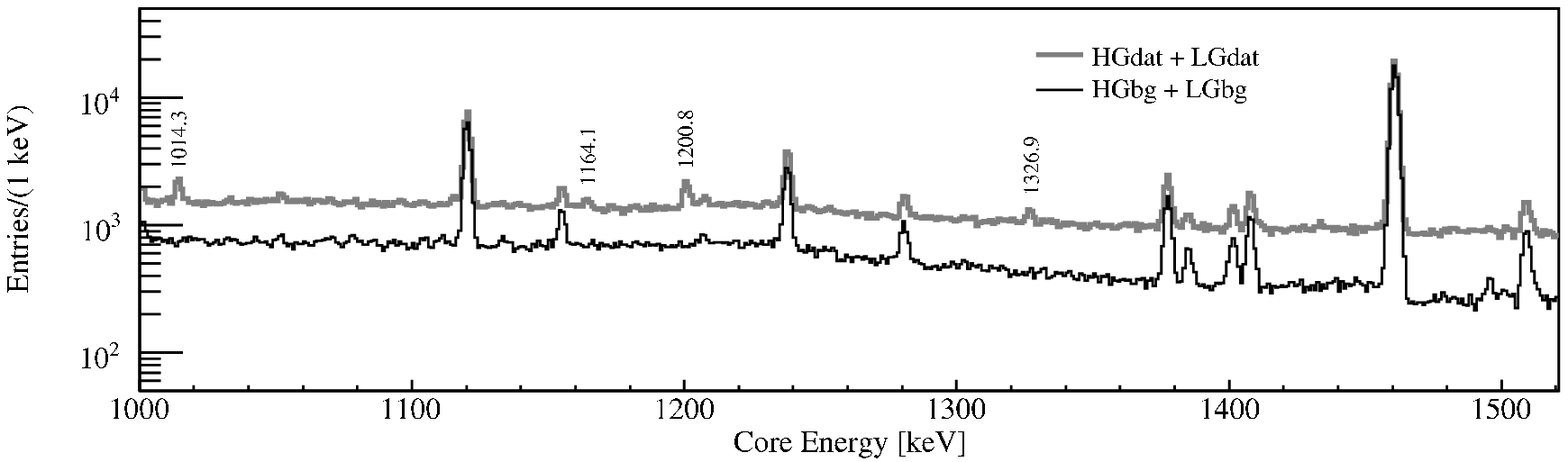}
  \includegraphics[width=\textwidth,clip]{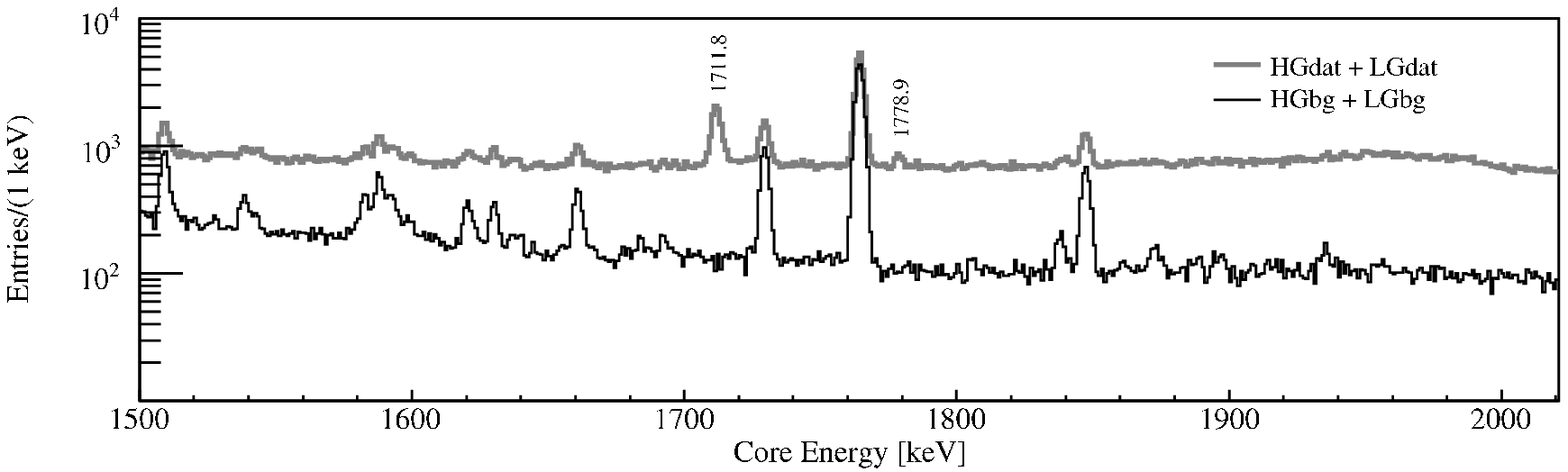}
  \includegraphics[width=\textwidth,clip]{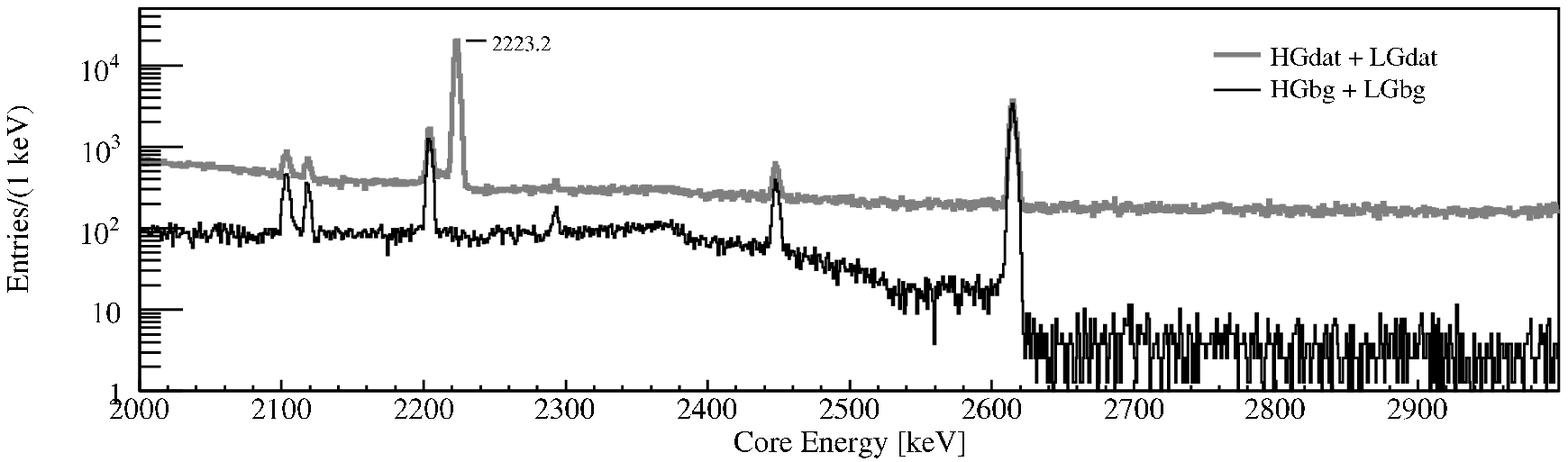}
  \caption{Core energy spectra with and without source. The
    normalization procedure is described in the text. The energy range
    is [0.08, 3]~MeV. Peaks induced by the AmBe source are indicated
    with their energies.}
  \label{fig:spec}
\end{figure}

The total energy deposited in the germanium crystal was read out from
the core electrode of the detector. Fig.~\ref{fig:spec} shows the core
energy spectra for the data and background in the range of [0.08,
3]~MeV. The thick line indicates the sum of HGdat and LGdat. The fine
line represents the normalized sum of HGbg and LGbg. The trigger
thresholds were set such that the spectra above 100~keV were not
affected.

Eight photon peaks from the background were fitted with a Gaussian
function plus a first order polynomial to normalize the background to
the data.  They are associated with the decays of $^{214}$Pb
(352~keV), $^{214}$Bi (609~keV, 1120~keV, 1764~keV, 2448~keV),
$^{228}$Ac (911~keV), $^{40}$K (1461~keV), and $^{208}$Tl
(2615~keV). The numbers of events in the peaks determined by the fits
were used to calculate the data to background ratios. The average of
these ratios, $1.279 \pm 0.003$, was then used to scale the background
spectrum.

\begin{figure}[tbhp]
  \centering
  \includegraphics[width=\textwidth,clip]{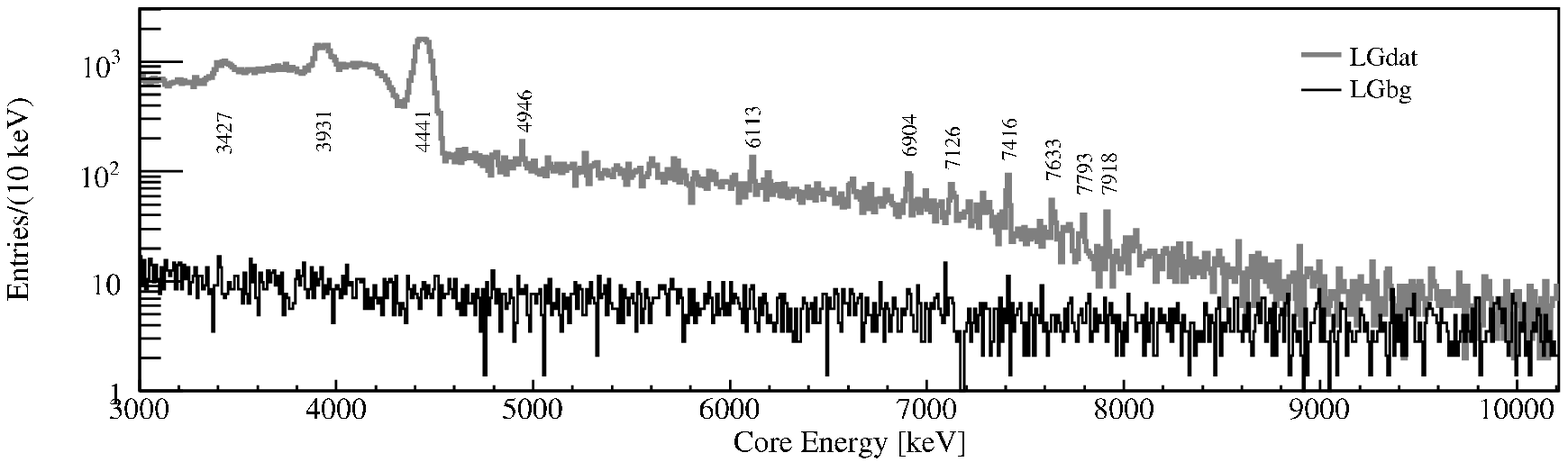}
  \caption{Core energy spectra with and without source. The
    normalization procedure is described in the text. The energy range
    is [3, 10.2]~MeV. Peaks induced by the AmBe source are indicated
    with their energies.}
  \label{fig:specl}
\end{figure}

Fig.~\ref{fig:specl} shows the spectra in the range of [3,
10.2]~MeV. In this energy range the background is small, as there are
hardly any natural radioactive elements producing photons with such
high energies.

\section{Neutron Interactions as Seen by the Core}
\label{sec:type}
The main interaction mechanisms of neutrons with energies less than
12~MeV are thermal capture, inelastic and elastic scattering. Elastic
scattering does not induce peaks that can be identified, because there
is no photon emitted and the recoil energy distribution is too flat.
Not only the production mechanism of the excited nucleus is important
for the identification of a peak. The de-excitation mechanism has also
to be taken into account. In most cases the nucleus de-excites
instantaneously with the emission of one or more photons. However, it
can also undergo internal conversion, in which case an electron from a
lower shell is emitted instead of a photon. The excited nucleus can
also be meta-stable and not de-excite instantaneously.

Table~\ref{tab:type} lists the processes identified in the core energy
spectrum. If inelastic scattering happens inside the germanium
crystal, the nuclear recoil energy is recorded as well as the energies
from some of the prompt photons. In case of instantaneous
de-excitation they are summed up: $E_{inelastic} = E_{\gamma} +
E_{recoil}$. This causes an asymmetric peak with a long recoil tail on
the high energy side. Internal conversions only create identifiable
peaks, if they occur inside the crystal, otherwise the emitted
electrons do not reach the detector.

\begin{table}[tbhp] 
  \caption{Type of neutron processes identified in the core energy spectrum.}
  \label{tab:type}
  \begin{tabular*}{\textwidth}{@{\extracolsep{\fill}}c|c|c|c}\hline\hline
    Production & De-excitation & Symbolic Notation & Short Form \\\hline
    & instantaneous & $n + ^A$Z$ \rightarrow ^{(A+1)}$Z$ + \gamma$ & $^A$Z$(n,\gamma)$\\\cline{2-4}
    thermal capture & meta-stable & $n + ^A$Z$ \rightarrow ^{(A+1)m}$Z, & $^A$Z$(n,\gamma^{m})$\\
    & & $^{(A+1)m}$Z$ \rightarrow ^{(A+1)}$Z$+\gamma$ &  \\\hline
    inelastic & instantaneous & 
    $n + ^A$Z$ \rightarrow ^A$Z$ + n^\prime + \gamma$ & $^A$Z$(n,n^\prime\gamma)$\\\cline{2-4}
    scattering & internal conversion & 
    $n + ^A$Ge$ \rightarrow ^A$Ge$^{+} + n^\prime + e^-$ & $^A$Ge$(n,n^\prime e)$\\\hline\hline
  \end{tabular*}
\end{table}

Table~\ref{tab:peak}, \ref{tab:peak2} list all the peaks observed in
the core energy spectrum due to neutron interactions within the
germanium crystal as well as within the surrounding materials: H, C,
Cl in the paraffin collimator, iron barrel of the collimator, Al, Ce
in the aluminum vacuum can, and copper container of the
pre-amplifiers.

Pure photon peaks were fitted with a Gaussian function plus a first
order polynomial to get the mean energies, FWHMs and the numbers of
events in the peaks. The 596~keV peak from $^{74}$Ge$(n, n^\prime
\gamma)$ does not have a Gaussian distribution. The treatment of this
peak will be described in section~\ref{sec:seg}. The 662~keV peak
associated with $^{140}$Ce has a significant background contribution
from $^{137}$Cs. This was subtracted. The 692~keV peak from
$^{72}$Ge$(n,n^\prime e)$ is hard to fit because it is asymmetric and
broad. It is also contaminated by other peaks nearby. The number of
events in this peak was estimated by integration.

The 4.4 MeV peak is due to photons from the de-excitation of
$^{12}$C$^{*}$ created in the AmBe source by the interaction,
$^{9}$Be$(\alpha,n)^{12}$C$^{*}$. It is Doppler broadened because of
the movement of the $^{12}$C$^{*}$ nuclei. The width of this peak
listed in Tab.~\ref{tab:peak2} was determined by the fit. Above 5 MeV
the widths of most of the peaks had to be fixed in the fitting
procedure due to low statistics. The peaks that are not identified are
marked with a question mark.

\begin{table}[tbhp] 
  \caption{Peaks observed in the core energy spectrum (see Fig.~\ref{fig:spec})
    due to neutron interactions.} 
  \label{tab:peak}
  \begin{tabular*}{\textwidth}{@{\extracolsep{\fill}}p{2.5cm}p{2.5cm}cc}\hline\hline
    Fitted Energy [keV]&Fitted FWHM [keV] &Interaction Type  &No. of Events\\\hline
    139.4       & $1.6 \pm 0.2$ & $^{74}$Ge$(n,\gamma^m)$        & $3377 \pm 520$ \\
    197.9       & $1.9 \pm 0.2$ & $^{70}$Ge$(n,\gamma^m)$        & $3306 \pm 503$ \\
    499.8       & $1.9 \pm 0.7$ & $^{70}$Ge$(n,\gamma)$          & $503  \pm 186$ \\
    595.7 *     & -             & $^{74}$Ge$(n,n^\prime\gamma)$   & $(18.4 \pm 2.5)\times10^3$\\
    662.0 $\dag$& $1.9 \pm 0.1$ & $^{140}$Ce$(n,\gamma)$         & $2802 \pm 188$ \\
    685.6       & $1.4 \pm 0.2$ & ?                             & $628  \pm 111$ \\
    692  $\ddag$& -             & $^{72}$Ge$(n,n^\prime e)$       & $\sim 7000$    \\
    708.5       & $2.4 \pm 0.5$ & $^{35}$Cl$(n,\gamma),^{36}$Cl$\rightarrow^{36}$Ar& $782  \pm 197$ \\
    721.9       & $1.9 \pm 0.2$ & ?                             & $3502 \pm 148$ \\
    843.4       & $2.4 \pm 0.5$ & $^{27}$Al$(n,n^\prime\gamma)$   & $1558 \pm 202$ \\
    846.6       & $2.4 \pm 0.2$ & $^{56}$Fe$(n,n^\prime\gamma)$   & $2802 \pm 196$ \\
    867.8       & $1.9 \pm 0.5$ & $^{73}$Ge$(n,\gamma)$          & $425  \pm 129$ \\
    962.2       & $2.4 \pm 0.2$ & $^{63}$Cu$(n,n^\prime\gamma)$   & $1041 \pm 129$ \\
    1014.3      & $2.4 \pm 0.2$ & $^{27}$Al$(n,n^\prime\gamma)$   & $1958 \pm 123$ \\
    1164.1      & $2.6 \pm 0.5$ & $^{35}$Cl$(n,\gamma)$          & $646  \pm 140$ \\
    1200.8      & $2.8 \pm 0.2$ & DEP of 2223 $\star$           & $2318 \pm 122$ \\
    1326.9      & $2.4 \pm 0.2$ & $^{63}$Cu$(n,n^\prime\gamma)$   & $711  \pm 91$  \\
    1711.8      & $3.8 \pm 0.1$ & SEP of 2223 $\star$           & $5555 \pm 133$ \\
    1778.9      & $2.6 \pm 0.2$ & $^{27}$Al$(n,\gamma),^{28}$Al$\rightarrow^{28}$Si & $469  \pm 73$ \\
    2223.2      & $3.8 \pm 0.1$ & $^{1}$H$(n,\gamma)$            & $79349 \pm 300$\\
    \hline\hline
  \end{tabular*}
  * The fitting of the 596~keV peak is described in a later section.\newline
  $\dag$ The background contribution to the 662~keV peak was subtracted.\newline
  $\ddag$ The number of events in the 692~keV peak was determined by integration.\newline
  $\star$ SEP, DEP stand for Single Escape Peak and Double Escape Peak, respectively. \newline
  ? The unidentified peaks are marked with a question mark.
\end{table}

\begin{table}[tbhp] 
  \caption{Peaks observed in the core energy spectrum (see Fig.~\ref{fig:specl})
    due to neutron interactions.} 
  \label{tab:peak2}
  \begin{tabular*}{\textwidth}{@{\extracolsep{\fill}}llcc}\hline\hline
    Fitted Energy [keV]&Fitted FWHM [keV]&Interaction Type&No. of Events\\\hline
    3427        & $85 \pm 7$    & DEP of 4441 $\star$            & $2354 \pm 263$ \\
    3931        & $87 \pm 5$    & SEP of 4441 $\star$            & $5873 \pm 368$ \\
    4441        & $92 \pm 2$    & $^{9}$Be$(\alpha,n)^{12}$C$^{*}$ & $14672 \pm 297$ \\
    4946        & $4.9\pm1.4$   & $^{12}$C$(n,\gamma)$            & $68 \pm 15$ \\
    6113        & 7 $\dag$      & $^{35}$Cl$(n,\gamma)$           & $75 \pm 12$ \\
    6904        & 7 $\dag$      & SEP of 7416 $\star$            & $60 \pm 10$ \\
    7126        & 7 $\dag$      & ?                              & $38 \pm  9$ \\
    7416        & 7 $\dag$      & $^{35}$Cl$(n,\gamma)$           & $70 \pm 10$ \\
    7633        & 7 $\dag$      & $^{56}$Fe$(n,\gamma)$           & $18 \pm 10$ \\
    7793        & $7.1\pm2.1$   & $^{35}$Cl$(n,\gamma)$           & $21 \pm  8$ \\
    7918        & $6.8\pm1.4$   & $^{63}$Cu$(n,\gamma)$           & $29 \pm  8$ \\
    \hline\hline
  \end{tabular*}
  $\dag$ The widths were fixed during the fit. \newline
  $\star$ SEP, DEP stand for Single Escape Peak and Double Escape Peak, respectively. \newline
  ? The unidentified peaks are marked with a question mark.
\end{table}

\section{Neutron Interactions as Seen by the Segments}
\label{sec:seg}
The energies deposited in each of the segments are read out
separately. This provides more information about the interactions
inside the germanium crystal than can be extracted from the core
signal alone.

For example, a photon with an energy of the order of one MeV has a
mean free path of several centimeters in the germanium crystal. It
most probably deposits energy in several different segments because of
multiple Compton scattering. The result is a \emph{multi-segment
  event}, in short MSE. In contrast, if there is only one segment with
an energy deposition, it is called a \emph{single-segment event}, in
short SSE. The power of discrimination of MSE and SSE induced by
photon using segmented germanium detectors has been shown in
\cite{pid}.

\subsection{Neutron Inelastic Scattering}
Compared to photon induced events, a special characteristic of neutron
inelastic scattering in a germanium crystal is that not only the
photon energy, but also the recoil energy, is recorded. The inelastic
scattering peak in the core energy spectrum has a high energy recoil
tail and, hence, is much less significant than a pure photon peak with
the same number of events, as seen in Fig.~\ref{fig:spec}. However, it
is possible to partially separate out the recoil energy distribution
using information from the individual segments. The disentangled
photon peak is much more significant than the original peak. This is
due to the way a segmented detector can provide information about
event topologies.

\begin{figure}[tbhp]
  \centering
  \includegraphics[width=0.9\textwidth]{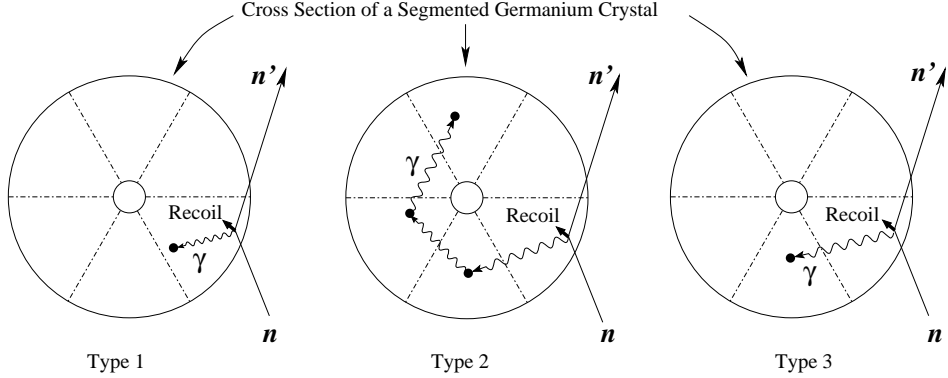}
  \caption{Three topologies of neutron inelastic scattering inside a
    germanium crystal.}
  \label{fig:inel}
\end{figure}

Figure~\ref{fig:inel} shows the three types of events contributing to
the inelastic scattering peak in the core spectrum. In all cases the
scattered neutron escapes:
\begin{enumerate}
\item The nuclear recoil energy and the prompt photon energy are
  deposited in the same segment;
\item The nuclear recoil energy is deposited within one segment, the
  prompt photon deposits its energy in several other segments;
\item The nuclear recoil energy is deposited within one segment while
  the prompt photon deposits its total energy within another segment.
\end{enumerate}

In the first case, only one segment has a signal. The energies
recorded by the core and the segment are the same, i.e. $E_{core} =
E_{seg} = E_{\gamma} + E_{recoil}$. Segmentation cannot help to
disentangle the two energies. In the second case, the recoil energy
can be observed in one segment. As the photon energy is shared between
several segments, there is no peaked distribution in any single
segment. This would partially be recovered by segment energy
summation. In the third case, the recoil energy is observed in one
segment, while the photon is observed in another segment. To
disentangle the photon peak from the recoil energy distribution,
energy spectra of all the segments are summed to get a spectrum of the
energy deposited in \emph{any segment}. In this spectrum, the type~1
events produce the same distribution as in the core spectrum and type
2 events form a flat distribution. The type~3 events, however, create
a sharp photon peak at the original energy and an enhancement in the
low energy region, the recoil energy distribution.

\begin{figure}[tbhp]
  \centering
  \includegraphics[width=\textwidth,clip]{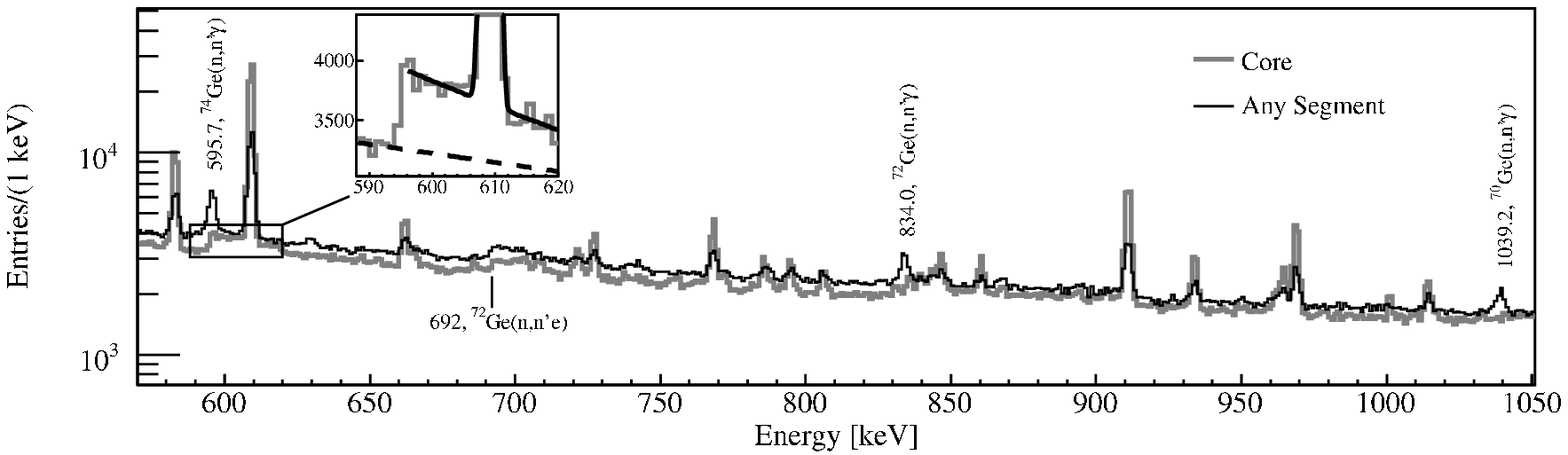}
  \includegraphics[width=\textwidth,clip]{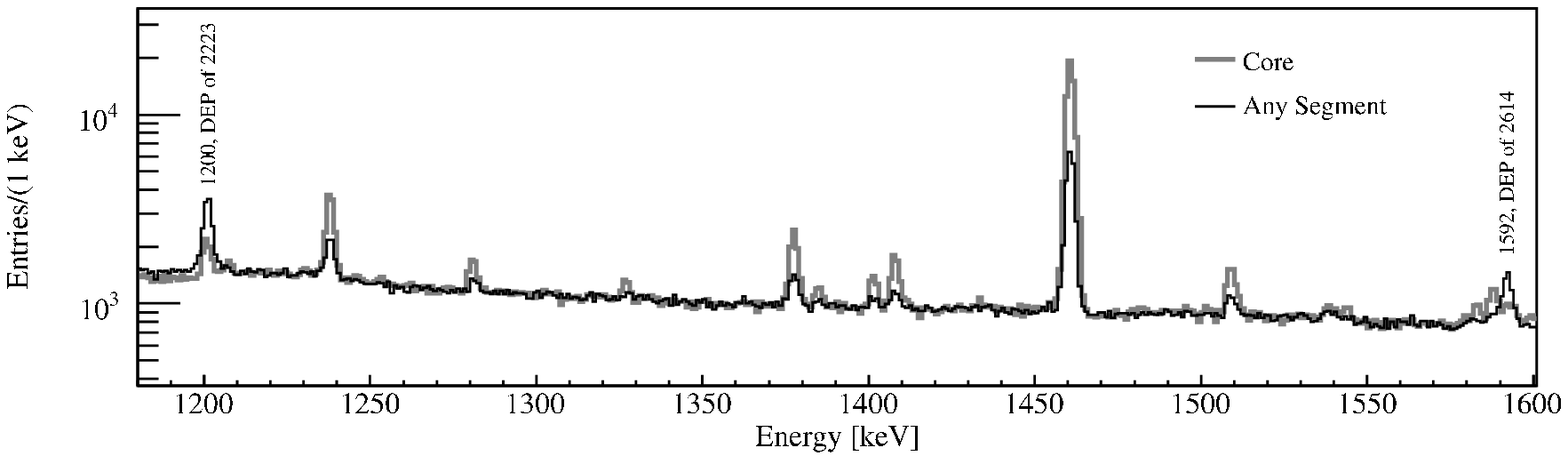}
  \caption{The \emph{any segment} spectrum (fine line) and the core
    spectrum (thick line) in the relevant energy ranges. The inset
    shows a close-up of the 596~keV peak in the core spectrum. The
    fitting of recoil energy distribution is described in the text.}
  \label{fig:cas}
\end{figure}

Figure \ref{fig:cas} shows the \emph{any segment} spectrum (fine line)
together with the core spectrum (thick line) in the relevant energy
ranges. Three peaks at 596~keV, 834~keV and 1039~keV, associated with
inelastic scattering, $^{74}$Ge$(n, n^\prime\gamma)$, $^{72}$Ge$(n,
n^\prime\gamma)$ and $^{70}$Ge$(n, n^\prime\gamma)$, respectively, are
clearly visible in the \emph{any segment} spectrum. The latter two are
washed out in the core spectrum.

It is possible to extract the number of each type of events in the
596~keV peak. The core spectrum was used to determine the total number
of events, $N_{total}$. An exponential function was fitted to the
shoulder of the peak associated with the nuclear recoil energy
distribution. A Gaussian function was fitted to the 609~keV background
photon peak on the shoulder simultaneously. The background below the
recoil structure was obtained from interpolating the spectrum below
and above the shoulder. The total number of events equals to the
difference between the fitted exponential and the background. The
error was estimated by assuming different levels and shapes of the
background. The fit is shown in the inset of Fig.~\ref{fig:cas}. The
dashed line represents the background, the solid line shows the
exponential plus the Gaussian function from the background peak.

The number of type~3 events, $N_{type3}$, was obtained by fitting a
Gaussian function plus a first order polynomial to the 596~keV peak in
the \emph{any segment} spectrum. The small shoulder caused by the
contamination with type~1 events does not change the results of the
fit significantly.

The determination of the number of type~1 events requires the study of
the peaks purely induced by photons. This provides the probability
that a de-excitation photon deposits its energy in exactly one or in
multiple segments. The relative strength of the peaks in the core and
any segment spectrum, that is, $\mathcal{R}(E_{\gamma}) =
N_{core}(E_{\gamma}) / N^{any}_{seg}(E_{\gamma}) = (N_{SSE} + N_{MSE})
/ N_{SSE}$, directly translates to the relative rate of the total
number of inelastic scattering events to the sum of type~1 and type~3
events, that is, $\mathcal{R}(E_{\gamma}^{inelastic}) = N_{total} /
(N_{type1} + N_{type3})$.

\begin{figure}[tbhp]
  \centering
  \includegraphics[width=0.5\textwidth,clip]{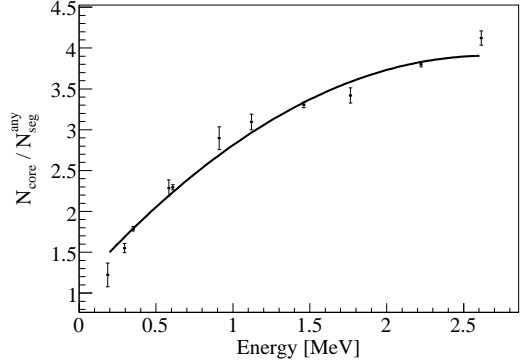}
  \caption{The ``core to any segment ratio'' as a function of the
    energy.}
  \label{fig:sf}
\end{figure}

A Gaussian function plus a first order polynomial were fitted to
eleven of the most prominent background photon induced peaks in the
core and \emph{any segment} spectra, respectively. The numbers of
events in the peaks from the fits were used to calculate the ratio,
$\mathcal{R}(E_{\gamma})$. The points with error bars in
Fig.~\ref{fig:sf} represent the ratios calculated at different
energies. A second order polynomial was fitted to get an estimate of
the ratio at any energy, $\mathcal{R}(E)$. The number of type~1 events
can be calculated as $N_{type1} = N_{total} /
\mathcal{R}(E_{\gamma}^{inelastic}) - N_{type3}$.

The number of type~2 events can then be calculated as $N_{type2} =
N_{total} - N_{type1} - N_{type3}$. The results concerning event
topologies in the 596~keV peak are listed in the second row of
Tab.~\ref{tab:ncore}. The percentage of single-segment events, that
is, $N_{type1}$, out of the total number of events is $\mathcal{P} =
N_{type1} / N_{total} \approx 5\%$.
\begin{table}[tbhp] 
  \caption{Numbers of events in the 596~keV, 834~keV and 1039~keV peaks.} 
  \label{tab:ncore}
  \begin{tabular*}{\textwidth}{@{\extracolsep{\fill}}lcccc}\hline\hline
    E [keV]& $N_{type 1}$ & $N_{type 2}$  & $N_{type 3}$    & $N_{total}$   \\\hline
    595.8  & $(1 \pm 1)\times10^3$ & $(10 \pm 3)\times10^3$ & $7285 \pm 218$ & $(18.4 \pm 2.5)\times10^3$ \\
    834.0  & [0, 380]    & [4100, 4700] & $2592 \pm 186$ & [6700, 7700] \\
    1039.2 & [0, 240]    & [2700, 3100] & $1429 \pm 182$ & [4100, 4800] \\
    \hline\hline
  \end{tabular*}
\end{table}

The numbers of type~3 events in the 834~keV and 1039~keV peaks were
obtained by fitting the \emph{any segment} spectrum. Since there is no
peak at these two energies in the core spectrum, it is impossible to
get $N_{total}$ from a fit. However, since the percentage $\mathcal{P}
= N_{type1} / N_{total}$ decreases with energy, $\mathcal{P}$ at
834~keV and 1039~keV should be less than
$\mathcal{P}$(596~keV). Taking into account the relation, $N_{type1} =
N_{total}/\mathcal{R} - N_{type3}$, the ranges of numbers of events in
different topologies in the 834~keV and 1039~keV peaks were
calculated. They are listed in Tab.~\ref{tab:ncore} as well.

The following steps were used to disentangle the recoil energy
$E_{recoil}$ spectrum of inelastic scattering with a prompt photon of
energy $E_\gamma$:
\begin{enumerate}
\item Exactly two segments having an energy deposition greater than
  10~keV were required.
\item If one segment had an energy deposition in the range
  $[E_\gamma-3\sigma, E_\gamma+3\sigma]$, where $\sigma$ was the
  detector energy resolution, the energy deposited in the other
  segment was used.
\end{enumerate}

The following steps produced the background to the recoil spectrum:
\begin{enumerate}
\item Exactly two segments having an energy deposition greater than
  10~keV were required.
\item Two energy side-bands, $[E_\gamma-6\sigma, E_\gamma-3\sigma]$
  and $[E_\gamma+3\sigma, E_\gamma+6\sigma]$ were defined.
\item If one segment had an energy deposition in the side-bands, the
  energy deposited in the other segment was used.
\end{enumerate}

\begin{figure}[tbhp]
  \centering
  \includegraphics[width=0.5\textwidth]{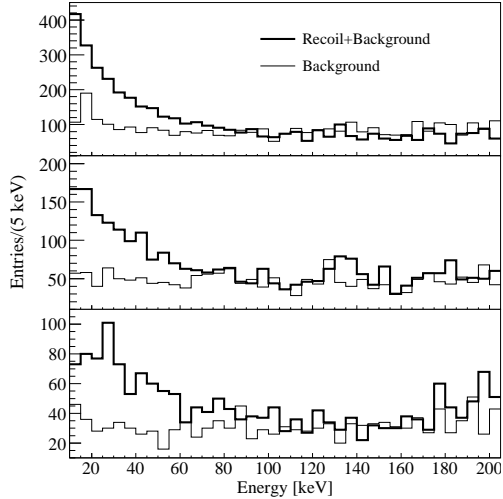}
  \caption{Recoil energy spectra corresponding to the inelastic
    neutron scattering with prompt photons of energies of 596~keV, 834
    ~keV and 1039~keV.}
  \label{fig:recoil}
\end{figure}

Fig.~\ref{fig:recoil} shows the disentangled recoil spectra related to
the 596~keV, 834~keV and 1039~keV photon peaks. The histograms start
at 10~keV. The spectrum is dominated by the electronic noise
below. The recoil spectra extending to $\sim 100$~keV are clearly
visible above the background.

\subsection{Internal Conversion}
\label{sec:conv}
If the excited state of a nucleus has the same spin as the ground
state, internal conversion~\cite{inco1, inco2} is the predominant mode
of the de-excitation. Since the mean free path of an electron emitted
from internal conversion is about 1~mm in germanium, the energy of the
electron and the recoil of the nucleus are deposited in the same
segment. The core and the \emph{any segment} spectra are the
same. This is demonstrated in Fig.~\ref{fig:cas}. The 692~keV peak
from internal conversion, $^{72}$Ge$(n,n'e)$, is neither changed nor
suppressed in the \emph{any segment} spectrum.

\subsection{Double Escape Peaks}
\label{sec:dep}
The double escape peaks are enhanced in the \emph{any segment}
spectrum, because many events from the single escape and full energy
peaks in the core spectrum move to this peak. Two enhanced double
escape peaks at 1200~keV and 1592~keV are clearly visible in
Fig.~\ref{fig:cas}. They originate from the 2223~keV peak of
$^{1}$H$(n,\gamma)$ and the 2614~keV peak of $^{208}$Tl.

\section{Verification of Simulations}
\label{sec:sim}
MaGe, a C++ simulation package developed by the Monte Carlo groups of
the Majorana and Gerda collaborations, was used to simulate the
experiment. It is based on Geant4~\cite{g1,g2}. The version Geant4 8.2
with patch-01 was used.

\subsection{Generator, Geometry and Process}
\label{sec:simdetail}
Figure~5 in~\cite{amben} shows the measured neutron spectrum emitted
from an AmBe source. It was normalized to a probability density
function and used in the neutron generator to assign energies to the
outgoing neutrons. The generator also produced 4.4 MeV photons from
the $^{12}$C$^{*}$ de-excitation inside the AmBe source. The Doppler
broadening of the 4.4 MeV peak was simulated by Gaussian smearing with
the observed widths taken from Tab.~\ref{tab:peak2}.

The geometry of the experiment was implemented according to technical
drawings. Approximations in the order of several centimeters had to be
made regarding
\begin{itemize}
\item the shape and size of the AmBe source and how it is held inside
  the paraffin collimator,
\item the exact relative position between the crystal and the paraffin
  collimator,
\item the exact geometry of the components inside the cryostat.
\end{itemize}

Geant4 provides high precision models for the simulation of
interactions of neutrons with energy below 20~MeV~\cite{g1,g2}. The
models depend on the ``evaluated neutron data library'' (G4NDL) for
cross sections, angular distributions and final state information. The
version G4NDL3.10 was used.

\subsection{Core Spectrum}
\label{sec:spemc}
\begin{figure}[tbhp]
  \centering
  \includegraphics[width=\textwidth,clip]{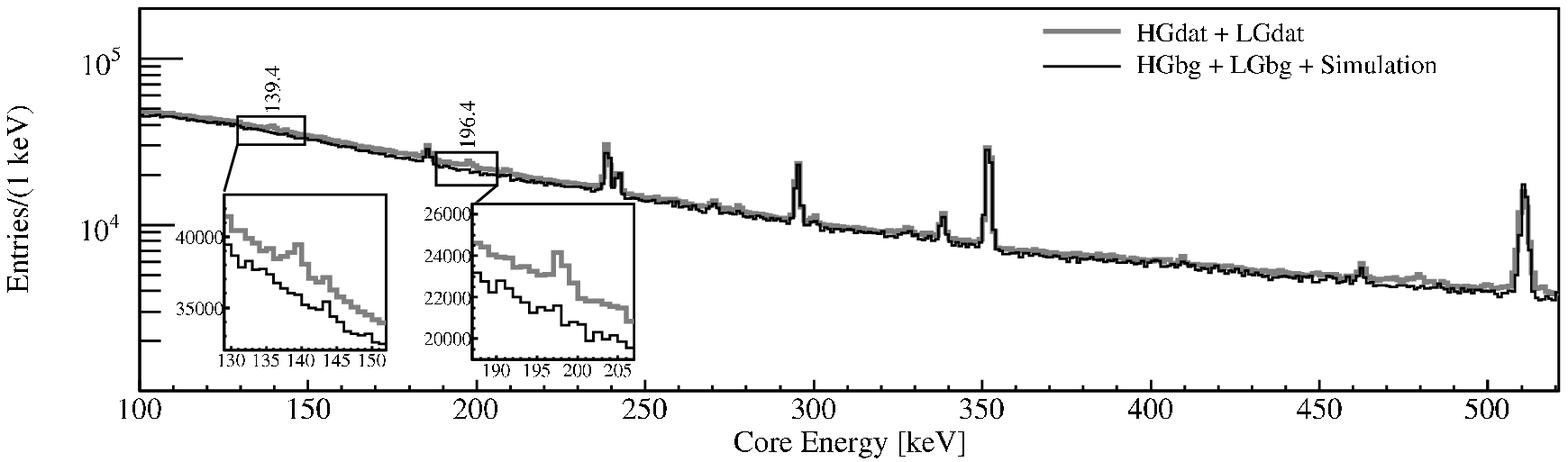}
  \includegraphics[width=\textwidth,clip]{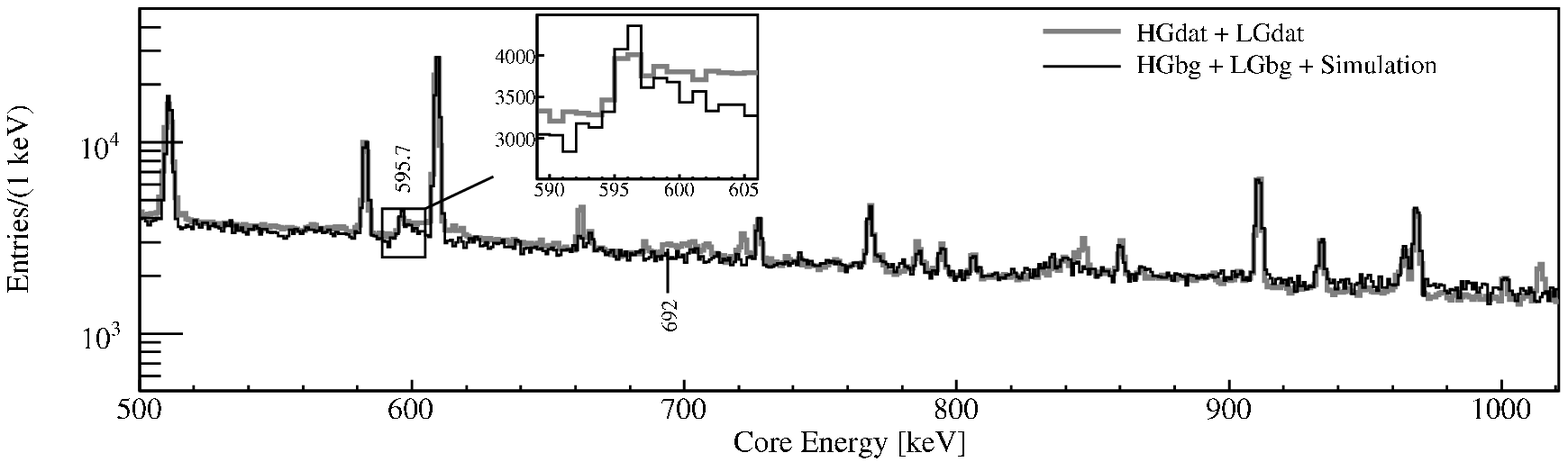}
  \includegraphics[width=\textwidth,clip]{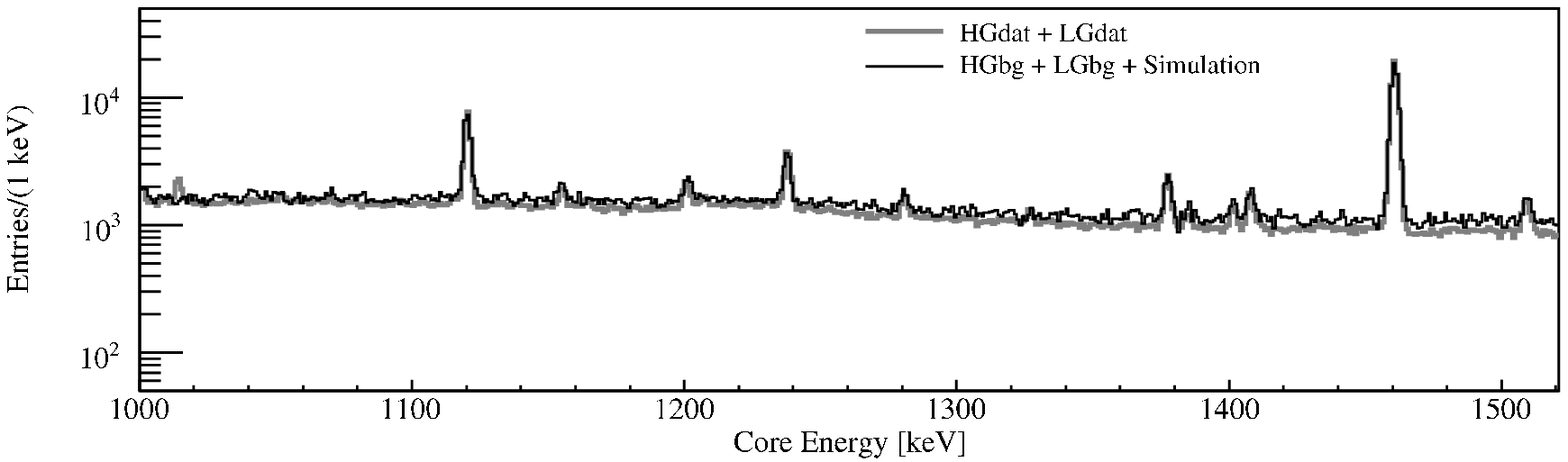}
  \includegraphics[width=\textwidth,clip]{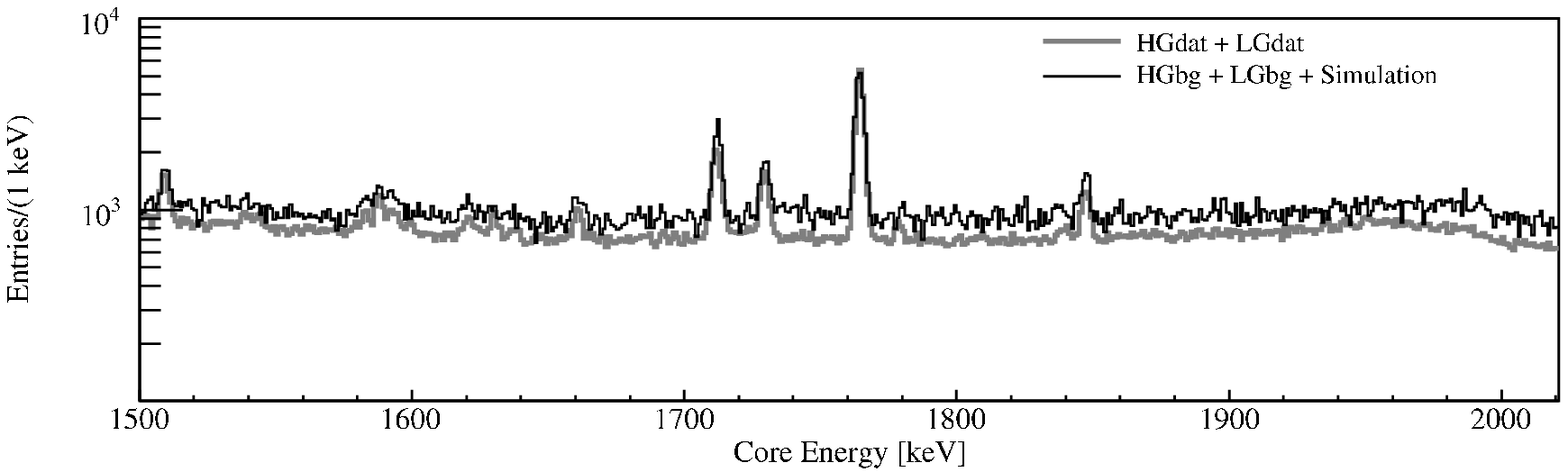}
  \includegraphics[width=\textwidth,clip]{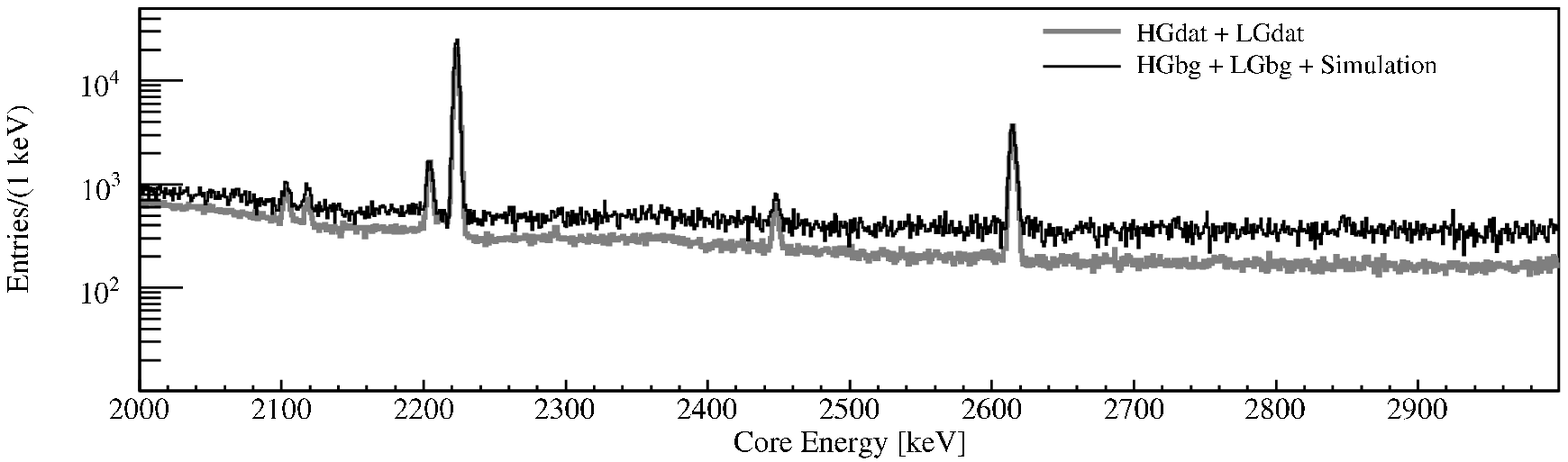}
  \caption{Comparison of the neutron core energy spectra from 0.1~MeV
    to 3~MeV between data and simulation plus measured background.}
  \label{fig:mc}
\end{figure}

\begin{figure}[tbhp]
  \centering
  \includegraphics[width=\textwidth,clip]{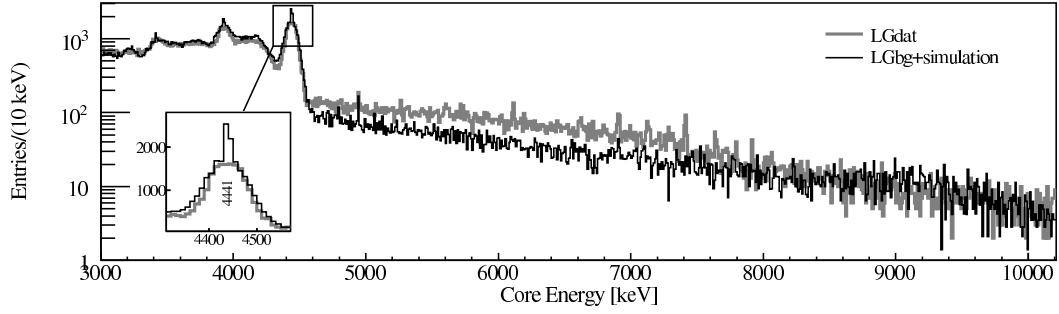}
  \caption{Comparison of the neutron core energy spectra from 3~MeV to
    10.2~MeV between data and simulation plus measured background.}
  \label{fig:mcl}
\end{figure}

Fig.~\ref{fig:mc} compares the core spectra of the measurement and the
simulation in the range of [0.1, 3]~MeV. The threshold effects below
100~keV were not taken into account in the simulation. The thick line
is the experimental data. The fine line is the sum of the simulation
and measured background. The background was normalized to data as
described in section \ref{sec:spec}. The simulation was normalized to
data according to the relation, $N_{data} = N_{background} +
N_{signal} = N_{background} + N_{simulation}$, where the $N$s are the
event numbers in the data, background and simulated
spectra. Fig.~\ref{fig:mcl} shows the same spectra in the range of
[3, 10.2]~MeV.

\subsection{Discrepancies between Data and Simulation}
\label{sec:dine}
The shapes of the continuous spectra from the simulation and data
deviate due to the poor knowledge of the exact material and geometry
of components between the source and the crystal.

There is a known bug~\cite{g4bug1} in Geant4 concerning neutron
inelastic scatterings. The secondary particles are not boosted back to
the laboratory frame after the calculations in the center of mass
frame are completed. This causes two problems:
\begin{itemize}
\item The simulated recoil energies of the germanium isotopes are
  wrong.
\item The photon peaks from the interactions are not broadened.
\end{itemize}

The first effect is demonstrated in the third inset of
Fig.~\ref{fig:mc}. The measured 596~keV peak from
$^{74}$Ge$(n,n^\prime\gamma)$ has a long tail on the high energy side
due to the nuclear recoil, while the simulated peak misses this
feature.

The second effect is demonstrated in the inset of
Fig.~\ref{fig:mcl}. The simulation generates a broad and a narrow
peak, both at 4.4 MeV. The broad peak is due to the de-excitation of
$^{12}$C$^{*}$ created in the source. The generator was adjusted
according to the data. The narrow one is due to neutron inelastic
scattering on carbon atoms in the paraffin collimator,
$^{12}$C$(n,n'\gamma)$. In reality, the carbon atom can gain a
velocity of up to $0.02c$ causing a Doppler broadening of the order of
50~keV-100~keV. This is comparable to the broadening in the
$^{12}$C$^{*}$ de-excitation peak, and can, thus, not be resolved in
the measured spectrum.

\begin{figure}[tbhp]
  \centering
  \includegraphics[width=0.5\textwidth]{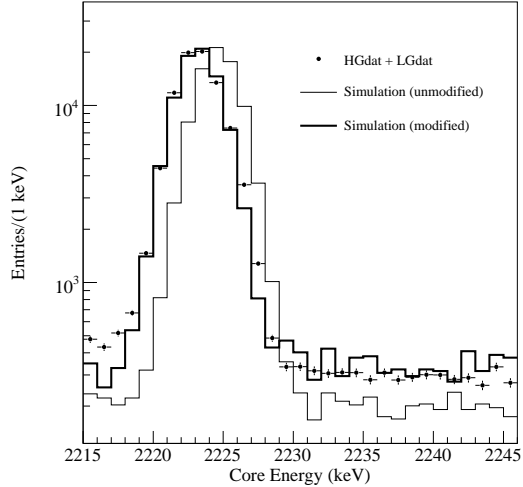}
  \caption{The 2223~keV photon peak from H$(n,\gamma)$ in data and
    simulation. The simulated peak is shifted at 2224.6~keV before the
    modification described in the text.}
  \label{fig:h2223}
\end{figure}

The mean value from the Gaussian fit to the measured 2223~keV photon
peak is $(2223.24 \pm 0.01)$~keV. The simulated peak centers at
$(2224.61 \pm 0.01)$~keV. This shifted value comes from the evaluated
neutron data library. This problem has been reported to the Geant4
Problem Tracking System~\cite{g4bug2}. It was fixed for our studies by
changing the value in the database to the measured energy. The result
is shown in Fig.~\ref{fig:h2223}.

The 139~keV and 196~keV photon peaks from the meta-stable states of
$^{75}$Ge and $^{71}$Ge produced by neutron captures are missing in
the simulated neutron spectrum, see the first two insets of
Fig.~\ref{fig:mc}. This problem has been reported to the Geant4
Problem Tracking System~\cite{g4bug3}.

The 692~keV peak from internal conversion, $^{72}$Ge$(n,n^{\prime}e)$,
is also missing in the simulation, see Fig.~\ref{fig:mc}. It also has
been reported~\cite{g4bug4}.

\section{Conclusion and Outlook}
\label{sec:out}
An 18-fold segmented germanium detector was exposed to an AmBe neutron
source and spectra were taken. A number of peaks from neutron
interactions on germanium isotopes as well as the surrounding
materials were identified. The segment information proved to be very
helpful in identifying these peaks. Inelastic neutron scattering
produces many events with energy depositions in more than one
segment. Hence, the improved understanding of neutron induced
interactions can also help to reduce the related background in the
$0\nu2\beta$ decay experiment, GERDA.

The Geant4 based simulation package, MaGe, was used to simulate the
experiment. Several discrepancies between data and MC were
found. Further verification and improvement of the related Geant4
codes are needed.

The experiment was not shielded from photons originating in the
source. This resulted in a low signal (neutron induced peaks) to
background (Compton shoulders of high energy photons) ratio. A new
experiment including photon shields will be performed to suppress the
photon induced background and further investigate the neutron
interactions on germanium isotopes.

\section{Acknowledgement}
\label{sec:ack}
The authors would like to thank the GERDA and Majorana Monte Carlo
groups for their fruitful collaboration and cooperation on the MaGe
project.




\begin{thebibliography}{00}




\bibitem{gerda} S.~Sch\"onert {\it et al.} [GERDA Collaboration],
  Nucl.\ Phys.\ Proc.\ Suppl.\ {\bf 145} (2005) 242.

\bibitem{pid} I. Abt {\it et al.}, arXiv:nucl-ex/0701005v1, and
  references therein.

\bibitem{siegfried} Iris Abt {\it et al.}, Nucl. Inst. and Meth. A
  {\bf 577} (2007) 574.

\bibitem{g1} S. Agostinelli {\it et al.}, [Geant4 Collaboration],
  Nucl. Inst. and Meth. A {\bf 506} (2003) 250.

\bibitem{g2} J. Allison {\it et al.}, IEEE Trans. Nucl. Sci. {\bf 53}
  (2006) 207.

\bibitem{mage} M.~Bauer {\it et al.}, Journal of Physics,
  Conf. Series.  {\bf 39} (2006) 362.

\bibitem{major} C. E. Aalseth {\it et al.}, [MAJORANA Collaboration],
  Nucl. Phys. B (Proc. Suppl.) {\bf 138} (2005) 217.

\bibitem{amben} J. W. Marsh {\it et all.}, Nucl. Inst. and Meth. A
  {\bf 366} (1995) 340.

\bibitem{geiger} K. W. Geiger, L. Van Der Zwan, Nucl. Inst. and Meth.
  {\bf 131} (1975) 315.

\bibitem{daq} Digital Gamma Finder (DGF) PIXIE-4, User’s Manual, X-Ray
  Instrumentation Associates, http://www.xia.com.


\bibitem{inco1} D. Lister, A. B. Smith, Phys. Rev. {\bf 183} (1969)
  954.

\bibitem{inco2} J. J. Kraushaar {\it et al.}, Phys. Rev. {\bf 101}
  (1956) 139.


\bibitem{ambeg} K. W. Geiger, L. Van Der Zwan, Nucl. Inst. and
  Meth. {\bf 131} (1975) 315.

\bibitem{g4bug1}\note Refer to ``problem 675'' in the Geant4 Problem
  Tracking System: http://bugzilla-geant4.kek.jp/.

\bibitem{g4bug2}\note Refer to ``problem 955'' in the Geant4 Problem
  Tracking System: http://bugzilla-geant4.kek.jp/.

\bibitem{g4bug3}\note Refer to ``problem 956'' in the Geant4 Problem
  Tracking System: http://bugzilla-geant4.kek.jp/.

\bibitem{g4bug4}\note Refer to ``problem 957'' in the Geant4 Problem
  Tracking System: http://bugzilla-geant4.kek.jp/.


\end{thebibliography}
\end{document}